\documentclass[sigconf]{acmart}

\usepackage[utf8]{inputenc}
\usepackage[draft]{minted}
\usepackage{geometry}

\usepackage{diagbox}
\usepackage{booktabs}
\usepackage{colortbl}
\usepackage{multirow}
\usepackage{tabularx}
\usepackage{soul}
\usepackage{wrapfig}
\usepackage[caption=false]{subfig}
\usepackage[font=footnotesize,labelfont=bf]{caption}

\usepackage{makecell}

\renewcommand\footnotetextcopyrightpermission[1]{} 
\setminted[verilog]{
    frame=lines,
    framesep=2mm,
    baselinestretch=1.1,
    fontsize=\footnotesize,
    linenos,
    breaklines
}

\AtBeginDocument{%
  \providecommand\BibTeX{{%
    \normalfont B\kern-0.5em{\scshape i\kern-0.25em b}\kern-0.8em\TeX}}}

\copyrightyear{2024}
\acmYear{2024}
\setcopyright{acmlicensed}\acmConference[GLSVLSI '24]{Great Lakes Symposium on VLSI 2024}{June 12--14, 2024}{Clearwater, FL, USA}
\acmBooktitle{Great Lakes Symposium on VLSI 2024 (GLSVLSI '24), June 12--14, 2024, Clearwater, FL, USA}
\acmDOI{10.1145/3649476.3660390}
\acmISBN{979-8-4007-0605-9/24/06}

\begin{document}

\title{Evolutionary Large Language Models for Hardware Security: \\A Comparative Survey}

\author{Mohammad Akyash}
\email{mohammad.akyash@ucf.edu}
\affiliation{
    \institution{ECE Department, University of Central Florida}
    \city{Orlando}
    \state{Florida}
    \country{USA}
}

\author{Hadi M Kamali}
\email{hadi.mardanikamali@ucf.edu}
\affiliation{
    \institution{ECE Department, University of Central Florida}
    \city{Orlando}
    \state{Florida}
    \country{USA}
}

\renewcommand{\shortauthors}{Mohammad Akyash and Hadi M Kamali}

\renewcommand{\shorttitle}{Evolutionary LLMs for Hardware Security: A Comparative Survey}

\begin{abstract}
    
Automating hardware (HW) security vulnerability detection and mitigation during the design phase is imperative for two reasons: (i) It must be before chip fabrication, as post-fabrication fixes can be costly or even impractical; (ii) The size and complexity of modern HW raise concerns about unknown vulnerabilities compromising CIA triad. While Large Language Models (LLMs) can revolutionize both HW design and testing processes, within the semiconductor context, LLMs can be harnessed to automatically rectify security-relevant vulnerabilities inherent in HW designs. This study explores the seeds of LLM integration in register transfer level (RTL) designs, focusing on their capacity for autonomously resolving security-related vulnerabilities. The analysis involves comparing methodologies, assessing scalability, interpretability, and identifying future research directions. Potential areas for exploration include developing specialized LLM architectures for HW security tasks and enhancing model performance with domain-specific knowledge, leading to reliable automated security measurement and risk mitigation associated with HW vulnerabilities.
    
\end{abstract}



\keywords{Large Language Models, Hardware Security, RTL Debugging}

\maketitle

\section{Introduction}

In today's semiconductor technology landscape, As system-on-chip (SoC) designs integrate more and more intellectual property (IP) cores, each with unique functionality and security challenges, each from various vendors, each with ever-increasing complexity, we witness a growing challenge in detecting and fixing vulnerabilities. Given the pivotal role of SoCs, while substantial efforts have been invested in software (SW) testing and debugging, SoC (HW-based) testing, validation, and verification remain less mature \cite{witharana2022survey}. The problem worsens while bugs are detected at lower levels of abstraction, which makes respins extremely difficult (and even impossible, e.g., post-silicon) \cite{azar2022fuzz}. Moreover, existing solutions, from simulation to formal verification, usually require expertise. Such solutions also suffer from scalability issues, unable to cope with the growing size and complexity of SoCs \cite{inamdar2021development}. Furthermore, these solutions cannot address the majority of SoCs' vulnerabilities due to rapidly evolving threats, such as zero-day attacks.

\begin{figure}[t]
    \centering
    {{\includegraphics[width=\columnwidth]{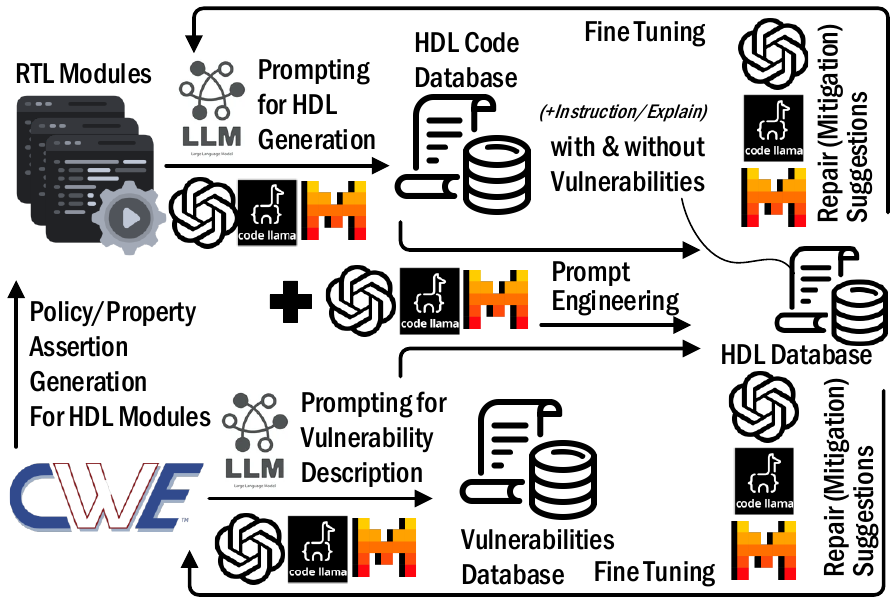}}}
    \caption{The Usage of LLMs for HDL (RTL) Generation/Validation.}
    \label{fig:llm_rtl_top}
\end{figure}

With the rapid evolution of LLMs, their capabilities have expanded into the domain of SW code generation with remarkable success, e.g., OpenAI's Codex \cite{chen2021evaluating}. Moreover, the scope of LLMs extends to SW code testing and verification while outperforming techniques like fuzzing \cite{liu2024your}. While significant progress has been achieved in SW through LLMs, studies at the HW/SoC level, particularly at RTL, have been dispersed. Many studies have initiated the LLMs' applicability at the HW/SoC level by raising questions like whether "LLM can generate HDL" or "LLM can validate HW designs". Just like in SW, LLMs have the potential to be utilized for both HW design, testing and validation (see Fig. \ref{fig:llm_rtl_top}). These studies show harnessing LLMs' capability to analyze, comprehend, and generate/validate complex code structures, might make them a right target vs. existing formal tools to identify potential security vulnerabilities within RTL codes \cite{ahmad2024hardware, kande2024security}. However, ensuring the integrity and security of HW designs, coupled with the potential for unknown vulnerabilities, presents broader challenges. 

This survey aims to offer a useful and comprehensive snapshot of rapidly growing use of LLMs in HW/SoC designs, particularly for security. We explore advancements, analyzing the pros and cons of each method. By examining current approaches, this work highlights the innovative application of LLMs to automate the detection and resolution of security vulnerabilities in HW designs. Also, we investigate future research directions, emphasizing the need for specialized LLM architectures and domain-specific knowledge integration. Our goal is to outline a roadmap for harnessing the full potential of LLMs in addressing HW security challenges, setting the stage for more robust and secure HW systems.

\section{LLMs for SW: Engineering and Testing}

Since the 1950s, many research efforts have been undertaken to develop highly efficient automated code generation tools \cite{gulwani2017program}. These efforts have spanned from traditional program synthesizers \cite{gulwani2017program}\footnote{Synthesizers aim to automatically generate programs (SW codes), based on a space search over a variety of constraints relevant to domains known as Domain Specific Languages (DSLs). These techniques are mostly limited to pre-defined DSLs and thus suffer scalability, being general-purpose, and adaptability issues \cite{desai2016program}.}, either deductive or inductive, to current neural-based models, notably codebase-reliant generative models \cite{austin2021program}. 

With recent outrageous advancements in LLMs, massive research has focused on applying LLMs for independent SW code generation, leading to widely-used platforms like Codex and CodeGen \cite{nijkamp2023codegen}. The foundation of these models lies in autonomously predicting the subsequent token by considering the preceding context, typically comprising function signatures and docstrings that describe the intended functionality of the program, translating human-written instructions into precise code snippets or entire programs \cite{nijkamp2023codegen}.

While this code generation relies on natural language processing (NLP), unlike natural language that is typically parsed as a sequential array of words or tokens, code generation is scrutinized based on its syntactic and semantic structure, often depicted using tree structures, e.g., abstract syntax trees (AST) \cite{ren2020codebleu}. Also, programming languages have a limited set of keywords, symbols, and rules, unlike the broad and nuanced vocabulary of natural languages. 

Given such differences, the primary concern for LLM-generated code is (i) correctness (testing and verification process), and (ii) codebase data hungriness \cite{ren2020codebleu}. In terms of correctness, testing and validation from the viewpoint of LLMs require well-defined metrics, where traditional metrics, e.g., \textit{BLEU} that widely used in NLP assessments \cite{ren2020codebleu}, fail due to their focus on linguistic similarity. For example, \textit{CodeBLEU} that evaluates the quality of code produced by LLMs, or \textit{Pass@k} that quantitatively measures the functional accuracy of code generation models, are example of such new metrics \cite{chen2021evaluating}. Regarding codebase data for code generation, substantial codebase data\footnote{The data must be not only vast but also diverse, relevant, and of high integrity as the superioir quality codebase data enhances model performance significantly \cite{liu2024your}.} is required for enhanced training and/or fine-tuning to improve the efficacy of LLMs for code ganeration \cite{chen2021evaluating, nijkamp2023codegen}. 

\section{LLMs for HW: Design and Testing}

Similar to SW engineering and testing, leveraging LLMs can significantly optimize and enhance circuit design processes, particularly within Electronic Design Automation (EDA) frameworks. LLMs can be used at high level abstraction, e.g., RTLs, to (i) reduce manual efforts for implementation\footnote{It can potentially serve as an alternative to high level synthesis (HLS), thereby enabling designers with limited HDL expertise to swiftly generate HW designs \cite{shi2023sechls}.}, (ii) address the challenge of lacking HDL codebase\footnote{Lack of HDL codebase is always a substantial barrier for AI-driven HW solutions, consequently enhancing the efficiency of the training phase \cite{azar2020nngsat}.}, (iii) expedite time-to-market (TTM) in the competitive chip design process, and (iv) enable a more efficient and reliable system (by reducing human-induced faults) \cite{shi2023sechls}. 

The current LLM-based methodologies in HW can be classified into two primary categories: (1) Development of automated AI agents aimed at streamlining EDA workflows (e.g., ASIC flow); (2) Derivation of SW code generation for RTL implementation. Regarding the former category, LLMs assist in various tasks such as script generation, architecture specification, and interpretation of compilation reports, thereby minimizing the workload of the design team. Within the latter category, solutions predominantly utilize LLMs in two manners: (i) \underline{refinement of design prompts}, which entails the creation (engineering) of more precise prompts to guide LLMs towards RTL generation with increased effectiveness, and (ii) \underline{RTL-based tuning}, which involves directly tuning LLMs through training on RTL code examples. A comparison of all existing LLM-based approaches in these two categories is shown in Table \ref{tab:rtl_eda_comparison}.

\begin{table*}
\fontsize{6pt}{6pt}\selectfont
\centering
\caption{A Top Comparison of LLM-based HW RTL Generation and EDA Tools.}
\label{tab:rtl_eda_comparison}
\setlength\tabcolsep{3pt} 
\begin{tabular}{@{} p{40pt} p{65pt} p{70pt} p{75pt} p{90pt} p{135pt}@{}}
\toprule  
Study & Target & LLM Engine & Input & Output & Comment (---Shortcomings---) \\
\cmidrule(r){1-1}\cmidrule(r){2-2}\cmidrule(r){3-3}\cmidrule(r){4-4}\cmidrule(r){5-5}\cmidrule(r){6-6}
Chang \emph{et al.} \cite{chang2023chipgpt} & RTL Generation + Refinement & GPT-3.5 & Design Specification Prompts \newline \textcolor{black}{+ Human Feedback for Corrections} & RTL Module & \textcolor{black}{- Static PPA analysis is post-LLM with no LLM-based improvement.} \newline \textcolor{black}{- Human feedback is needed for manual correction per design.}\\
\cmidrule(r){1-1}\cmidrule(r){2-2}\cmidrule(r){3-3}\cmidrule(r){4-4}\cmidrule(r){5-5}\cmidrule(r){6-6}
Thakur \emph{et al.} \cite{thakur2023autochip} & RTL Generation \newline \textcolor{black}{w/ guaranteed Compilation} & GPT-4, Llama2, \textcolor{black}{GPT-3.5T, Claude 2}	& Design prompt + Compile/Synthesis Report & Compiled and Tested RTL Design & \textcolor{black}{- Feedback addresses compilation/simulation errors but may alter function priority, leading to unintended functions.} \newline \textcolor{black}{- No Feedback for PPA Efficiency Matter}\\ 
\cmidrule(r){1-1}\cmidrule(r){2-2}\cmidrule(r){3-3}\cmidrule(r){4-4}\cmidrule(r){5-5}\cmidrule(r){6-6}
He \emph{et al.} \cite{he2024chateda} & Automatic EDA Flow Scripting and Execution Calls & 	Llama2-70B	& Natural Language Instructions \textcolor{black}{+ RTL Design}	& EDA Tool \textcolor{black}{Commands \& }Reports + \textcolor{black}{Scripts} + Synthesized Design + Layout (GDSII) & - It is either design- or technology-Dependent. \newline - Cannot be easily design/tool-agnostic. \\
\cmidrule(r){1-1}\cmidrule(r){2-2}\cmidrule(r){3-3}\cmidrule(r){4-4}\cmidrule(r){5-5}\cmidrule(r){6-6}
Li \emph{et al.} \cite{li2024specllm} & Architecture Specifications Generation + Review & GPT-4 & Architecture specifications + RTL Design	& Hierarchical Reviewed Architecture Specifications & \textcolor{black}{- Specifications are limited to the existing technologies.} \newline \textcolor{black}{- It is mostly processor-based instructions. Not for generic HW.}\\
\cmidrule(r){1-1}\cmidrule(r){2-2}\cmidrule(r){3-3}\cmidrule(r){4-4}\cmidrule(r){5-5}\cmidrule(r){6-6}
Lu \emph{et al.} \cite{lu2023rtllm} & RTL Generation & GPT-3.5, GPT-4, VeriGen, StarCoder & Natural language instructions	& RTL Design & - WIth no feedback, success rate is low for functional correctness. \newline
- The reference designs are very limited and relatively small.\\
\cmidrule(r){1-1}\cmidrule(r){2-2}\cmidrule(r){3-3}\cmidrule(r){4-4}\cmidrule(r){5-5}\cmidrule(r){6-6}
Liu \emph{et al.} \cite{liu2024rtlcoder} & RTL Generation & RTLCoder & Natural language instructions & RTL Design &  \textcolor{black}{- Diversity rate is low in the training dataset.} \newline \textcolor{black}{- The functional correctness of training dataset is not ensured, leading to lower functional coverage in the generated outputs.} \\
\cmidrule(r){1-1}\cmidrule(r){2-2}\cmidrule(r){3-3}\cmidrule(r){4-4}\cmidrule(r){5-5}\cmidrule(r){6-6}
Thakur \emph{et al.} \cite{verigen}  & Completing Partial RTL Design & MegatronLM-355M, CodeGen, code-davinci-002, and J1-Large-7B & Partial RTL Design + Custom problem set with testbenches & RTL Design & - Lack of Organized Dataset. \newline - RTLLM shows the performance does not surpass existing commercial models. \newline \textcolor{black}{- Completion necessarily does not provide correct functionalities.}\\
\cmidrule(r){1-1}\cmidrule(r){2-2}\cmidrule(r){3-3}\cmidrule(r){4-4}\cmidrule(r){5-5}\cmidrule(r){6-6}
Cheng \emph{et al.} \cite{chang2024data} & RTL Generation + Repair + EDA Script Generation & Llama2-7B, Llama2-13B & Natural language descriptions + Verilog files + EDA scripts & Corrected Verilog code + Verilog code from descriptions + EDA scripts & \textcolor{black}{- For refinement, it is for syntactic errors (compilation issues).} \\
\cmidrule(r){1-1}\cmidrule(r){2-2}\cmidrule(r){3-3}\cmidrule(r){4-4}\cmidrule(r){5-5}\cmidrule(r){6-6}
DeLo \emph{et al.} \cite{delorenzo2024make} & RTL Generation & VeriGen-2B & \textcolor{black}{Natural language instruction + RTL modules description} & Compiled, Tested, and PPA Improved RTL Design & \textcolor{black}{- Tested on Small Toy Circuits, e.g., adders and MAC units.} \newline \textcolor{black}{- Stochastic behavior of MCTS. Less Improvement in More Iterations.} \\
\cmidrule(r){1-1}\cmidrule(r){2-2}\cmidrule(r){3-3}\cmidrule(r){4-4}\cmidrule(r){5-5}\cmidrule(r){6-6}
Li \emph{et al.} \cite{li2024circuit} & RTL Synthesis (Mapping) &  Circuit Transformer & Gate-Level Design (AIG) & Design Model (Truth Table) + Synthesized AIG & - Low Accuracy for Larger Circuits. \newline - Low Performance with no MCTS (Low Scalability). \\
\bottomrule
\end{tabular}
\vspace{-10pt}
\end{table*}

\subsection{LLM Agent for EDA Automation}

Several studies have explored the potential of LLM in automating the ASIC design/implementation process \cite{he2024chateda, li2024specllm, liu2023chipnemo, Blocklove_2023}. ChatEDA and ChipNeMo are two examples of task planning and execution agents that interpret natural language commands from the design team. ChipNeMo \cite{liu2023chipnemo} implements a series of domain-specific training strategies for chip design tasks. It involves the deployment of bespoke tokenizers, domain-adaptive continued pretraining, and supervised fine-tuning guided by domain-specific instructions. ChatEDA \cite{he2024chateda} aims to facilitate optimal interaction with the EDA tools by comprehending instructions in natural language for generating and delivering executable programs.

Using such techniques, LLM agents can offer automated ASIC flow, from RTL generation to GDSII creation, by invoking necessary SW tools and utilizing required scripts/files. However, while promising, these techniques necessitate thorough analysis to truly enhance automation in EDA tools for the following reasons:

\noindent \underline{(1) Expert-Oriented Training and Fine-Tuning}: Constructing such frameworks heavily relies on expert efforts for training or fine-tuning them to accommodate specific ASIC flows. Given the variety of technologies with their respective documentation, syntaxes, flows, and scripting methods, the pre-trained LLM may not offer a universally applicable model for all environments.

\noindent \underline{(2) Failure in Handling Unforeseen Incidents}: Despite extensive fine-tuning, the LLM-based agent may inaccurately extract information from reports/specs or generate incorrect scripts/configs when confronted with new incidents in the flows. Technology advancements, EDA tools updates, etc., may worsen this issue, as the LLM agent may fail to provide the desired output under evolving conditions.

\noindent \underline{(3) Dependence on Technology}: To clarify this, we raise a question! How similar is the EDA flow (i) from one design to another design, (ii) from one technology to another technology, (iii) from one vendor to another vendor? Now, the question becomes how deep is LLM fine-tuned based on these designs, technologies, and vendors? While chatbots may offer basic assistance, the prospect of achieving comprehensive automation seems to remain elusive.

\subsection{LLM for RTL Generation and Refinement}

The main LLM-based RTL-oriented research focuses on the generation and refinement of RTL, primarily transitioning from specification to RTL design (+optimization). Initial efforts emphasize prompt engineering, crucial to successful RTL generation while relying on the existing LLMs \cite{Blocklove_2023, chang2023chipgpt, lu2023rtllm}. Other methods, e.g., Verigen and VerilogEval, adapt open-source LLMs like CodeGen \cite{nijkamp2023codegen}, followed by fine tuning on RTL, to produce more optimized HDL modules \cite{verigen, liu2023verilogeval}. Additionally, studies such as ChipGPT and AutoChip explore use of feedback mechanisms to enhance HDL quality, addressing aspects like compilation errors and design optimization (PPA optimization) \cite{chang2023chipgpt, thakur2023autochip}. While these methods often rely on static analysis, DeLorenzo et al. Introduce optimization techniques like Monte Carlo tree search (MCTS) to fine-tune LLM tokens even further for more tuned optimization at the backend of LLMs \cite{delorenzo2024make}. 

More recent advancements have shifted the focus from fine tuning and prompt engineering in existing LLMs to the development of dedicated circuit transformers, e.g., Li et al. Introduce "Circuit Transformer" with 88M parameters and integrated MCTS for optimization, leading to a fully open-source independent LLMs for RTL \cite{li2024circuit}. Similarly, RTLCoder proposes an automated data generation flow utilizing a model with 7B parameters, producing a sizable labeled dataset for RTL generation \cite{liu2024rtlcoder}. These endeavors have led to the emergence of large circuit models (LCM), enhancing the expression of circuit data's semantics and structures, thus creating more robust, efficient, and innovative design approaches. 

Despite its promise, more research is needed as follows:

\noindent \underline{(1) Universality Issues}: LLM-based RTL generation faces limitations due to scarce codebase knowledge available for model fine-tuning and training per application \cite{liu2024rtlcoder}. As an example, developing security enclaves or fully-debugged Verilog modules is incredibly challenging as there are not many training datasets available for it.

\noindent \underline{(2) Verification (Functional) Issues}: Existing studies highlight the complex nature of (functional) verification tasks, further magnified by the limited availability of trained models for test bench generation and functional simulation \cite{liu2023verilogeval}. The complexity of circuit designs, which involve both functional and structural attributes, worsens the challenge, as even small changes to the structure (a code line) can have significant effects on functionality, underscoring the complexity of testbench generation and simulation of circuits.

\noindent \underline{(3) Scalability Issues}: Scalability is crucial for RTL-based LLMs in addressing complex circuit designs \cite{lu2023rtllm}. Efforts to enhance computational efficiency and model architecture sophistication are essential to accommodate larger designs and meet evolving electronic device demands. Further research is necessary to overcome scalability challenges and maximize LLM potential in RTL generation.

\section{LLM for HW: Security (Verification)}

Given the paramount significance of security of HW designs in modern SoCs, and in light of the earlier discussion emphasizing the importance of verification over LLMs, several studies have commenced employing LLM for SoC verification (moving towards bug-free designs, either functional or security-oriented). Similar to LLM-based RTL design, these approaches fall into two main categories: (i) \underline{refinement of design prompts}, where designers guide LLMs toward generating secure code (i.e. prompt engineering), and (ii) \underline{RTL-based tuning}, which is about altering the LLM's framework itself to generate output bug-free code. In advancing HW security, researchers have leveraged LLMs using either pure natural language prompts (i.e. description of the code) or a blend of natural language (i.e. comments designed by human experts) and code. The following describes these two categories in detail and how each category can enhance verification and security for HW designs.

\subsection{Prompt Engineering}

\begin{figure}[t]
    \centering
    {{\includegraphics[width=0.85\columnwidth]{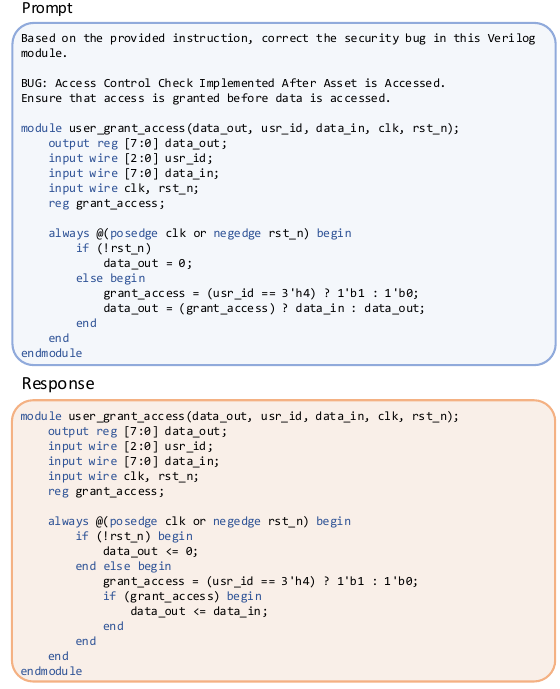}}}
    \caption{An Examplary Case in GPT-4 for Security Debugging.}
    \label{fig:prompt_fixing}
\end{figure}

Prompt engineering is the practice of designing inputs for LLMs, to obtain specific, desirable outputs. This technique optimizes the interaction with LLMs to improve its performance on various tasks, leveraging strategies like few-shot \cite{brown2020few}, and chain-of-thought \cite{wei2022chain} prompting to guide the model's responses effectively. A few recent studies in HW explore the applications of prompt engineering for enhancing vulnerability detection and repair, as well as design verification. For example, \cite{ahmad2024hardware} employs a range of detailed instruction prompts for various LLMs, aiming to evaluate the efficacy of each model in correcting HW vulnerabilities\footnote{These prompts must provide a thorough description of the bug, strategies for debugging, and illustrative examples that contrast insecure code with its secure counterpart.}. Fig. \ref{fig:prompt_fixing} shows an example of how prompting GPT-4 with a bug description and repair instructions alongside the Verilog code enables GPT-4 to address the vulnerability. Here are two important lessons to be learned:

\noindent (1) The example shows that being super specific is crucial in engineering the prompt to ensure the generated code is devoid of vulnerabilities. Thus, it is vital to have careful crafting by human experts to generate such prompts. This requirement for human input could become a tedious process, posing challenges in scaling and automating the approach for broader applications.

\noindent (2) The performance and efficacy of LLMs depends on the infrastructure of LLM used. While commercial LLMs like GPT-4 tend to outperform models trained on coding datasets, including Codegen and VeriGen, in terms of repair accuracy and efficacy, this advantage comes at the cost of increased number of parameters.

The importance of precision in prompt generation is also shown in \cite{nair2023generating}, relying on ChatGPT, revealing the fact that the success rate can be degraded significantly while the model is more limited\footnote{The number of parameters was restricted to a range of millions instead of billions.}. This study also demonstrates models misguiding the designers while the Verilog code of various CWE scenarios as part of instruction can lead to new form of vulnerabilities from prompts (may not fully represent the capture of potential vulnerabilities in SoC designs).  

To enhance verification capability, some studies focus on the use of LLMs for verification assertion generation (e.g., SystemVerilog Assertions (SVAs)). For instance, \cite{orenesvera2023using} uses GPT-4 in an iterative mechanism to refine prompts for GPT-4, enabling it to generate more accurate and complete SVA properties from RTL code. This approach coupled with AutoSVA2, which automatically generates formal verification testbenches, enables LLM-guided formal verification towards more automation. However, the major obstacle to this automation is the reliance of this approach on iterative refinement by an expert, which requires a deep understanding of both HW verification and prompt engineering.

Similarly, AssertLLM \cite{fang2024assertllm} uses a customized GPT-4 Turbo to generate SVAs (functional verification assertions) from natural language design specifications (translating design documents). Although results show high success rate, this model is also heavily dependent to the quality and completeness of the design documents. This is while richness of documentation is always a critical issue in HW design, thus AssertLLM might struggle to generate assertions that fully capture the intended design behavior.

LLM4DV \cite{zhang2023llm4dv} uses LLMs with prompt templates to automate the generation of test stimuli for verification. LLM4DV integrates LLMs with a systematic method that includes a stimulus generation agent, prompt templates, and four LLM-based improvements, e.g., summarizing prompts, resetting, etc. Evaluated using three custom-designed large-scale DUTs, this framework demonstrated promising results and achieved high coverage rates in simple scenarios. However, this approach focuses more on coverage-related metrics, overlooking security-oriented vulnerabilities.

Similar to these formal-based mechanisms, \cite{kande2024security} proposes designing an evaluation framework that includes generating natural language prompts that mimic code comments in assertion files, using these prompts to generate SVAs with LLMs, and then assessing the correctness of these assertions against a benchmark suite of real-world HW designs and corresponding golden reference assertions. The results demonstrate that LLMs, with varying levels of detail in the prompts, can generate valid HW security assertions.

More recent use of LLMs for RTL debugging aimed to enhance automation in the domain. For instance, RTLFixer \cite{tsai2024rtlfixer} automatically rectifies syntax errors in Verilog code by leveraging Retrieval-Augmented Generation (RAG) and the ReAct prompting strategy. RTLFixer employs a retrieval database filled with expert knowledge of syntax errors. ReAct also introduces an iterative approach involving reasoning, action, and observation, mimicking experts' debugging techniques. This combination builds a more effective system for automating the debugging. However, it still heavily relies on the comprehensiveness and currentness of the external knowledge database, which is collected by human experts.

\begin{table*}[t]
\fontsize{5.7pt}{5.7pt}\selectfont
\centering
\caption{A Top Comparison of LLM-based HW Security Validation Solutions}
\label{tab:existing_comparison}
\setlength\tabcolsep{3pt} 
\begin{tabular}{@{} p{25pt} p{40pt} p{40pt} p{25pt} p{40pt} p{40pt} p{80pt} p{55pt} p{110pt}@{}}
\toprule  
Study & Target & LLM Engine & \# of Bugs & Success Rate & Source of Benchmarks & Expert Knowledge Needed? & Reference (for Eval) & Comment \\
\cmidrule(r){1-1}\cmidrule(r){2-2}\cmidrule(r){3-3}\cmidrule(r){4-4}\cmidrule(r){5-5}\cmidrule(r){6-6}\cmidrule(r){7-7}\cmidrule(r){8-8}\cmidrule(r){9-9}
Nair \emph{et al.} \cite{nair2023generating} & \textcolor{black}{Prompt generation for Debugging RTL} & ChatGPT & 10 & 100\%$^{*1}$ & CWE (Descriptions)
 & \textcolor{black}{For the Whole Process} & Manual expert intervention per debugging & - Cannot be automated. \newline - Limited evaluation on CWEs \\
\cmidrule(r){1-1}\cmidrule(r){2-2}\cmidrule(r){3-3}\cmidrule(r){4-4}\cmidrule(r){5-5}\cmidrule(r){6-6}\cmidrule(r){7-7}\cmidrule(r){8-8}\cmidrule(r){9-9}
Kande \emph{et al.} \cite{kande2024security} & Detection (Generate Assertion) & OpenAI Codex (code-davinci-002)
& 10 & \textcolor{black}{$\sim$25\%} & Hack$@$DAC21, \newline OpenTitan & \textcolor{black}{For manually building detailed security constraints} & Golden Assertion & \textcolor{black}{- High success rate only when bug and security policy is known. Otherwise, it is below 10\%.} \newline - Only for single \texttt{endmodule}, No Hierarchical and Recursive SVA. \\
\cmidrule(r){1-1}\cmidrule(r){2-2}\cmidrule(r){3-3}\cmidrule(r){4-4}\cmidrule(r){5-5}\cmidrule(r){6-6}\cmidrule(r){7-7}\cmidrule(r){8-8}\cmidrule(r){9-9}
Ahmad \emph{et al.} \cite{ahmad2024hardware} & Repair \newline \textcolor{black}{(pre-detected bugs)} & OpenAI Codex (code-davinci-001, code-davinci-002, code-cushman-001), CodeGen
 & 15 & $\sim$31\% & CWE (Benchmark), \newline OpenTitan, \newline Hack$@$DAC21 & \textcolor{black}{- For training (dataset generation for assisting repairs)} \newline \textcolor{black}{- For CWEAT static analyze verification}  & Repaired Code (Prompt Reference) & - Only applicable on pre-observed cases with high similarity (to be detected by CWEAT) \\
\cmidrule(r){1-1}\cmidrule(r){2-2}\cmidrule(r){3-3}\cmidrule(r){4-4}\cmidrule(r){5-5}\cmidrule(r){6-6}\cmidrule(r){7-7}\cmidrule(r){8-8}\cmidrule(r){9-9}
Saha \emph{et al.} \cite{saha2023llm} & Detection (Generate Assertion), security vulnerability insertion & GPT 3.5, \newline GPT 4 & \textcolor{black}{N/R$^{*2}$} & \textcolor{black}{N/R$^{*2}$} & CWE, Trust-Hub & For prompt engineering and evaluation & Manual expert intervention per debugging & \textcolor{black}{- Limited evaluation on CWEs and smart toy circuits.}\\
\cmidrule(r){1-1}\cmidrule(r){2-2}\cmidrule(r){3-3}\cmidrule(r){4-4}\cmidrule(r){5-5}\cmidrule(r){6-6}\cmidrule(r){7-7}\cmidrule(r){8-8}\cmidrule(r){9-9}
Fu \emph{et al.} \cite{Fu_2023} & Detection and/or Repair & StableLM, Falcon, LLama2 & 1 \textcolor{black}{(different models)} & $\sim$35\% & Open-Source SoCs and Microprocessors & For fine-tuning (Open-source code classifications) & Repaired Code (Pre- and Post-correction of Git (CVA6, Opentitan, …)) & - Detailed enhancement for training is needed. Per design, a new training might be required. \newline - Raw dataset is limited and not design-agnostic). \\
\cmidrule(r){1-1}\cmidrule(r){2-2}\cmidrule(r){3-3}\cmidrule(r){4-4}\cmidrule(r){5-5}\cmidrule(r){6-6}\cmidrule(r){7-7}\cmidrule(r){8-8}\cmidrule(r){9-9}
Meng \emph{et al.} \cite{meng2023unlocking} & Detection (Generate Assertion) & HS-BERT & 8 & 326 Bugs from 1723 sentences & RISC-V, OpenRISC, MIPS, OpenSPARC, OpenTitan (documentation) & For classifying security rules in documents & \textcolor{black}{Manual expert labling for security property validation} & 
- Limited by the quality of the input HW documentation. \newline \textcolor{black}{- Limited to the design/verification team knowledge.}\\
\cmidrule(r){1-1}\cmidrule(r){2-2}\cmidrule(r){3-3}\cmidrule(r){4-4}\cmidrule(r){5-5}\cmidrule(r){6-6}\cmidrule(r){7-7}\cmidrule(r){8-8}\cmidrule(r){9-9}
Fang \emph{et al.} \cite{fang2024assertllm} & Detection (Generate Assertion) & GPT4 Turbo & N/A &  89\% & Open-source CPUs, SoCs, Xbars, arithmetic. & \textcolor{black}{For extracting verification-required information from  documents} & Golden RTL Implementation & - Limited by the quality of the input HW documentation. \newline \textcolor{black}{- Mostly syntactic and basic functional verification.}\\
\cmidrule(r){1-1}\cmidrule(r){2-2}\cmidrule(r){3-3}\cmidrule(r){4-4}\cmidrule(r){5-5}\cmidrule(r){6-6}\cmidrule(r){7-7}\cmidrule(r){8-8}\cmidrule(r){9-9}
Paria \emph{et al.} \cite{paria2023divas} & Detection (Generate Assertion) & ChatGPT, \newline BART & N/A & N/A & CEP SoC (MIT-LL) & \textcolor{black}{For assumptions (CWE-based security rules)} & N/R$^{*2}$ & - Expert review for Spec Generation is needed per design.\\
\cmidrule(r){1-1}\cmidrule(r){2-2}\cmidrule(r){3-3}\cmidrule(r){4-4}\cmidrule(r){5-5}\cmidrule(r){6-6}\cmidrule(r){7-7}\cmidrule(r){8-8}\cmidrule(r){9-9}
Vera \emph{et al.} \cite{orenesvera2023using} & Detection (Generate Assertion) & GPT-4 & \textcolor{black}{N/R$^{*2}$} & \textcolor{black}{N/R$^{*2}$} & RISC-V CVA6 & \textcolor{black}{For building rules related to assertions} & Previously developed formal tools (AutoSVA) & - The success rate heavily depends on expert's input for prompt engineering.\\
\cmidrule(r){1-1}\cmidrule(r){2-2}\cmidrule(r){3-3}\cmidrule(r){4-4}\cmidrule(r){5-5}\cmidrule(r){6-6}\cmidrule(r){7-7}\cmidrule(r){8-8}\cmidrule(r){9-9}
Zhang \emph{et al.} \cite{zhang2023llm4dv} & Test Stimuli Generation & GPT-3.5-turbo & N/A & 	small: $\sim$98\%, \newline large: $\sim$65\% & Self-designed RTL Designs & For prompts generation & Coverage Monitoring & \textcolor{black}{- Not for security purposes. Coverage-based testing.}\\
\cmidrule(r){1-1}\cmidrule(r){2-2}\cmidrule(r){3-3}\cmidrule(r){4-4}\cmidrule(r){5-5}\cmidrule(r){6-6}\cmidrule(r){7-7}\cmidrule(r){8-8}\cmidrule(r){9-9}
Tsai \emph{et al.} \cite{tsai2024rtlfixer} & Syntax Errors Repair & GPT-3.5, \newline GPT-4 & 212 & 98.5\% & VerilogEval benchmarks, \newline RTLLM benchmarks & For retrieval database (debugging reference) & VerilogEval, \newline RTLLM & \textcolor{black}{- Not for security purposes. Only for Syntax errors.} \\
\bottomrule
\multicolumn{7}{l}{$^{*1}$: It is 100\% as all the debugging is done manually. Bug is known, the debugging instruction (flow) is known, and GPT is used for generation.} & \multicolumn{2}{l}{N/R$^{*2}$: Not Reported.} \\
\end{tabular}
\vspace{-10pt}
\end{table*}

Some LLM-based studies focus on the use of such models at the SoC level. DIVAS \cite{paria2023divas} uses LLMs to analyze SoC specifications and crafts precise queries that encapsulate potential security vulnerabilities related to the SoC. These queries are submitted to LLMs, e.g., ChatGPT and Google's BARD, and the LLMs map these queries to relevant CWE vulnerabilities that could compromise the SoC. Once CWEs have been identified, DIVAS utilizes LLMs to construct SVAs for each. These SVAs are designed to act as security verification mechanisms, ensuring the SoC's design complies with security standards and is safeguarded against identified vulnerabilities.

Similarly, \cite{saha2023llm} explores how GPTs are utilized in SoC level for security vulnerability insertion, detection, assessment, and mitigation. This study, focusing on smaller models, e.g., ChatGPT-3.5, and relying on a sub-set of CWEs, evaluates the modification possibility over RTL using one- and few-shot learning. By comprehensive exploration, the study suggests specific prompt guidelines for effectively using LLMs in SoC security-related tasks.

LLMs possess a dual-use nature; While advancing HW security initiatives, LLM can also present new threats simultaneously. \cite{kokolakis2024harnessing} delves into the potential of general-purpose models like ChatGPT in the offensive HW security domain This study involves employing prompt engineering techniques to guide LLMs in filtering complex HW design databases, correlating system-level concepts with specific HW modules, identifying security-critical design modules, and modifying them to introduce HW Trojans. This study initiates the possibility of using LLMs for building more stealthy and undetectable HW Trojans, reshaping the characteristics of HW Trojan implementation, detection, and mitigation.

\subsection{Fine-Tuning}

As mentioned previously, some of these LLM-based HW verification solutions rely on fine-tuning, which involves adjusting a pre-trained language model by training it on Verilog/SVA data. However, LLMs require extensive datasets for effective training, posing a significant challenge in specialized domains, particularly in HW security due to the scarcity of targeted data. LLM4SecHW \cite{Fu_2023} is one example, which leverages a dataset compiled from defects and remediation steps in open-source HW designs, using version control data from GitHub. This dataset was created by selecting significant HW projects such as CVA6, CVA5, OpenTitan, etc., and extracting commits, issues, and pull requests (PRs) related to HW designs. This approach provides a rich source of domain-specific data for training models, specifically tailored to identifying and fixing bugs in HW designs. Although innovative and promising, the quality of this data is dependent on the filtering process accuracy. The effectiveness of LLMs in debugging HW designs is thus directly tied to how precisely the data is curated and processed.  

The NSPG framework \cite{meng2023unlocking} is another example of LLM solution for HW verification that offers a novel methodology for automating the generation of HW security properties utilizing fine-tuned LLMs. This approach is anchored by the development of a specialized language model for HW security, HS-BERT, which is trained on domain-specific data. Through deep evaluation on previously unseen design documents from OpenTitan, NSPG has proven its capability by extracting and validating security properties, showing security vulnerabilities within the OpenTitan design. However, a notable limitation of not only NSPG, but also all HW-oriented fine-tuned model for now lies in its dependency on the quality and scope of the HW documentation provided as input (which is almost super limited). As in the realm of HW/SoC design, this documentation often remains incomplete, inconsistent, or lack necessary detail, the precision and efficacy of the solution could be adversely affected.

\section{Takeaways and Future Directions}

In all facets of using LLMs for HW security, it becomes apparent that a significant hurdle, whether in HW design or in testing/verification, whether stemming from prompt engineering or fine-tuning, lies in the procurement and effective utilization of quality data \cite{gunasekar2023textbooks}. Also, as depicted in Table \ref{tab:existing_comparison}, creating specialized LLMs (e.g., LCMs) or employing pre-existing ones necessitates a deep expert knowledge to achieve a high success rate for generation, detection, and mitigation. Considering these two obstacles, despite being promising, the endeavor requires rigorous effort across multiple facets.

Creating a standard database reference is crucial for both training and evaluating the methods proposed in this domain. It facilitates a fair comparison among different techniques, ensuring that the pros/cons of each approach can be accurately assessed. Moreover, high-quality RTL data is indispensable for the optimal training of LLMs. It enables these models to learn the intricacies of RTL designs effectively, thereby enhancing their efficiency in security tasks.

Given the distinct characteristics of RTL codes as opposed to natural language texts, it becomes crucial to consider domain-specific models for handling HW codes. Incorporating concepts such as graphs and ASTs into LLMs can bridge the gap between the structural nuances of RTL codes and the inherently sequential processing of conventional language models. It is crucial to devise a novel metric specifically for evaluating the security coverage of RTL code examined by LLMs. This metric would serve as a critical feedback mechanism for LLMs, enabling them to assess and refine their output continually. By quantitatively measuring the security of RTL designs, the metric would allow LLMs to optimize their learning process towards generating code that is not only functionally correct but also adheres to high security standards.

Building on the foundational strategies mentioned above, further refinement can be achieved through the optimization of continuous prompts\footnote{For instance, the Prefix-Tuning concept  \cite{li2021prefix} involves the addition of trainable tokens to prompts, thus enabling more task-specific model responses.}. Such strategies also open the doors for mechanisms to enhance prompt automation for LLMs, e.g., auto-prompting\footnote{Auto prompting could significantly mitigate the automation challenge and enhance the feasibility of (secure) code (RTL) generation \cite{yin2023dynosaur}.}. These optimizations are open research directions potentially presenting a more feasible and efficient alternative to LLM fine-tuning.

\section{Conclusion}

This paper examined the use of LLMs in detecting/addressing security flaws in HW designs. We specifically analyzed their incorporation into RTL, revealing their independent problem-solving abilities in this domain. Our examination of existing approaches highlights both their benefits and drawbacks, notably scalability and accuracy issues. Also, we identified potential areas for future research. Our suggestion involves developing dedicated LLM architectures and datasets focused on HW security, indicating a path toward targeted improvements that could mitigate HW vulnerabilities.

\bibliographystyle{ACM-Reference-Format}
\bibliography{refs}


\end{document}